\documentclass[runningheads]{llncs}
\usepackage[T1]{fontenc}
\usepackage{graphicx}
\usepackage{color}

\begin{document}
\title{Machine Learning for Automated Mitral Regurgitation Detection from Cardiac Imaging}

\titlerunning{Automated MR Detection from Cardiac Imaging}

\author{Ke Xiao \inst{1} \and
Erik Learned-Miller \inst{1} \and
Evangelos Kalogerakis\inst{1} \and
James Priest \inst{2} \and
Madalina Fiterau \inst{1}}
%
\authorrunning{K. Xiao et al.}
%
\institute{University of Massachusetts Amherst, Amherst MA 01003, USA \\
\email{\{kexiao,elm,kalo,mfiterau\}@cs.umass.edu} \\
\and
Stanford University, Stanford CA 94305, USA \\
\email{jpriest@stanford.edu}}

\maketitle              
\begin{abstract}
Mitral regurgitation (MR) is a heart valve disease with potentially fatal consequences that can only be forestalled through timely diagnosis and treatment. Traditional diagnosis methods are expensive, labor-intensive and require clinical expertise, posing a barrier to screening for MR. To overcome this impediment, we propose a new semi-supervised model for MR classification called CUSSP. CUSSP operates on cardiac imaging slices of the 4-chamber view of the heart. It uses standard computer vision techniques and contrastive models to learn from large amounts of unlabeled data, in conjunction with specialized classifiers to establish \emph{the first ever automated MR classification system}. Evaluated on a test set of 179 labeled -- 154 non-MR and 25 MR -- sequences, CUSSP attains an F1 score of 0.69 and a ROC-AUC score of 0.88, setting the first benchmark result for this new task.

\end{abstract}

\section{Introduction}

\begin{figure}[!ht]
      \minipage{0.33\textwidth}
        \includegraphics[width=\textwidth]{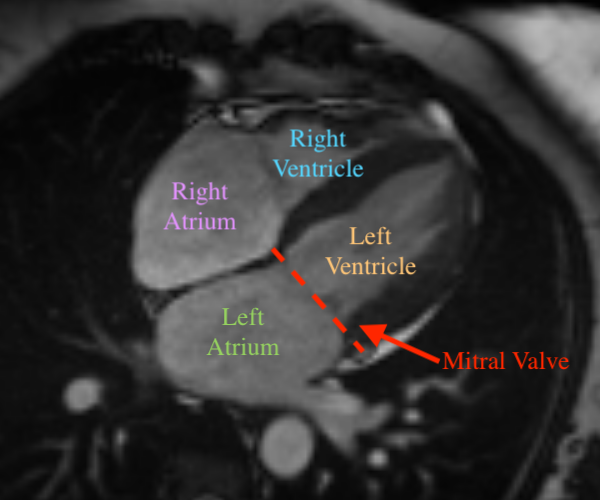}
      \endminipage\hfill
      \minipage{0.33\textwidth}
        \includegraphics[width=\textwidth]{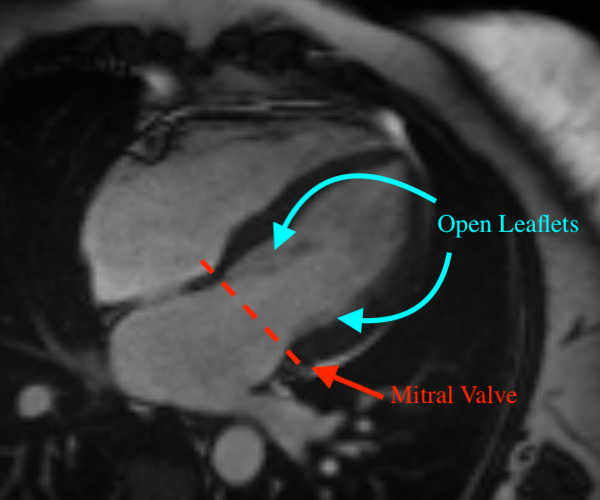}
      \endminipage\hfill
      \minipage{0.33\textwidth}
        \includegraphics[width=\textwidth]{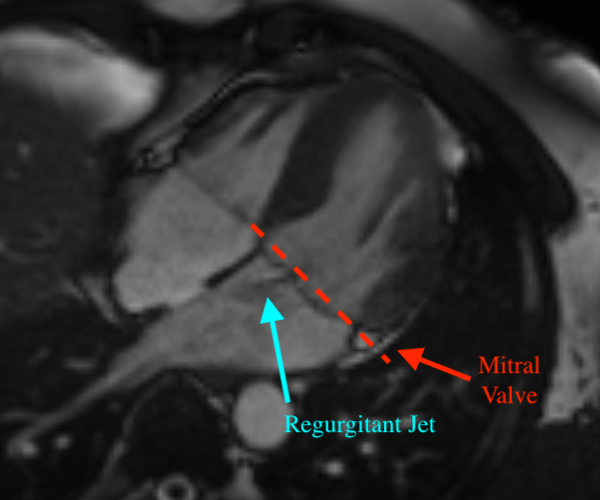}
      \endminipage\hfill
      \caption{Three cardiovascular magnetic resonance (CMR) images showing the long-axis four-chamber view of the heart. Left: a heart with normal mitral valve. Middle: a heart with normal mitral valve when the valve leaflets are open. Right: a heart with mitral regurgitation. The red dotted line denotes the mitral valve.}
      \label{fig:example_mri_4ch_denoted}
    \end{figure}

\paragraph{Mitral regurgitation.} Mitral regurgitation (MR)~\cite{enriquez2009mitral} is a valvular heart disease in which the mitral valve does not close completely during systole when the left ventricle contracts, causing regurgitation -- leaking of blood backwards -- from the left ventricle (LV), through the mitral valve, into the left atrium (LA) -- Figure~\ref{fig:example_mri_4ch_denoted}. 
MR can be caused by either organic or functional mechanisms~\cite{dziadzko2019causes}, with organic MR leading to atrial and annular enlargement and functional MR increasing atrial pressure. As MR progresses, it may cause arrhythmia, shortness of breath, heart palpitations and pulmonary hypertension~\cite{mirabel2007characteristics}. Left undiagnosed and untreated, MR may cause significant hemodynamic instability and congestive heart failure which can lead to death~\cite{parcha2020mortality}, while acute MR usually necessitates immediate medical intervention~\cite{watanabe2019acute}. Thus, \emph{early detection and assessment of MR are crucial for optimal treatment outcomes}, with the best short-term and long-term results obtained in asymptomatic patients operated on in advanced repair centers with low operative mortality ($<1\%$) and high repair rates ($\geq 80-90\%$)~\cite{enriquez2009mitral}.

\paragraph{MR diagnosis.} MR is often only detected following symptom onset. Among patients with asymptomatic MR, quantitative grading of mitral regurgitation is a powerful indicator for clinical treatment such as immediate cardiac surgery~\cite{enriquez2005quantitative}. Clinically, MR is usually diagnosed with doppler echocardiography, with \emph{cardiovascular magnetic resonance} (CMR) subsequently used to assess the MR severity and to accurately quantify the regurgitant volume, one of the indicators of severity~\cite{uretsky2018use}. Most studies that have evaluated CMR for assessing the mitral regurgitant volume use the difference between left ventricular stroke volume (LVSV) and forward stroke volume (FSV). LVSV is usually estimated with the short-axis (SA) view CMR -- a 4-D tensor -- while FSV is most commonly determined by aortic phase-contrast velocity-encoding images~\cite{uretsky2018use}. This diagnosis and assessment process requires continuous involvement from expert clinicians along with specific order and post-processing for the phase-contrast images of the proximal aorta or main pulmonary artery during the acquisition of the CMR data. The associated expense with this standard diagnostic procedure thus poses an obstacle to the large-scale screening for MR in the general population.

\paragraph{Toward machine learning for MR diagnosis.} Although quantitatively assessing mitral regurgitant volume requires specific CMR imaging sequences and expert analysis, four-chamber (4CH) CMR images provide a comprehensive view of all four heart chambers, including the mitral valve as it opens and closes, as shown in Figure~\ref{fig:example_mri_4ch_denoted}. Thus, we propose to train a model that uses 4CH CMR to automatically diagnose MR, making wide screening possible. As training data, we use the long axis 4CH CMR imaging data from the UK Biobank~\cite{doi:10.1126/scitranslmed.3008601}, from over 30,000 subjects, out of which N=704 were labeled by an expert cardiologist. While the 4CH view has the potential to identify MR when the regurgitant jet is visible, the imaging is not accompanied by comprehensive annotations or diagnoses of diseases/conditions for individual patients. To overcome this difficulty, we rely on weakly supervised and unsupervised methods. Weakly supervised deep learning has proved successful in detecting other heart pathologies. Specifically, Fries et al.~\cite{fries2019weakly} proposed a weakly supervised deep learning method (CNN-LSTM) to classify aortic valve malformation from the aortic valve cross section CMR present in the UK Biobank, wherein the critical feature of the aortic valve opening shape was easily extracted from the aortic valve cross section CMR imaging data. Meanwhile, Vimalesvaran et al.~\cite{vimalesvaran2022detecting} proposed a deep learning based pipeline for detecting aortic valve pathology using 3CH CMR imaging from three hospitals. The data set was fully annotated with landmarks, stenotic jets and regurgitant jets. Unlike these prior two studies, we faced the challenge of extracting complex mitral valve regurgitant features from 4CH CMR images with \emph{no annotations for landmarks, regurgitant jets or easily extractable features}, and only a small amount of binary MR labels. To the best of our knowledge, this is \emph{the first study on identifying MR using the 4CH CMR imaging data in an automated pipeline}.


\paragraph{Our approach.} We propose an automated five stage pipeline named \textbf{C}ardio-vascular magnetic resonance \textbf{U}-Net localized \textbf{S}elf-\textbf{S}upervised \textbf{P}redictor (CUSSP). Our approach incorporates several different preexisting neural network architectures in the pipeline, discussed in Section~\ref{sec:cussp}, to address the challenges inherent to the MR classification task. Specifically, we use a U-Net~\cite{ronneberger2015u} to perform segmentation of the heart chambers, which we then use to localize the area around the mitral valve. We apply histogram equalization to enhance the appearance of the valve. We then use a Barlow Twins~\cite{zbontar2021barlow} 
 network to learn, without supervision, representations of the blood flow around the valve, and a Siamese network~\cite{xing2018offline} to learn differences between instances of MR and non-MR. During training, CUSSP leverages a large amount of unlabeled CMR images, and \emph{minimal supervision}, in the form of a comparatively small set of MR labels manually annotated by cardiologist. However, \emph{at test time CUSSP is fully automated}.


\paragraph{Contribution.} Our work is the \emph{first study on automated detection of mitral regurgitation (MR)}, providing a benchmark for the classification of MR in an automated pipeline from long axis 4CH CMR images. Used as a screening tool, it has the potential to support hospital diagnostics and improve patient care.

\section{Methods}

\subsection{Segmentation of the cardiac magnetic resonance images}
\label{sec:segmentation}

\begin{figure}[!ht]
    \vspace{-4ex}
      \minipage{0.33\textwidth}
        \includegraphics[width=\textwidth, height=0.7\textwidth]{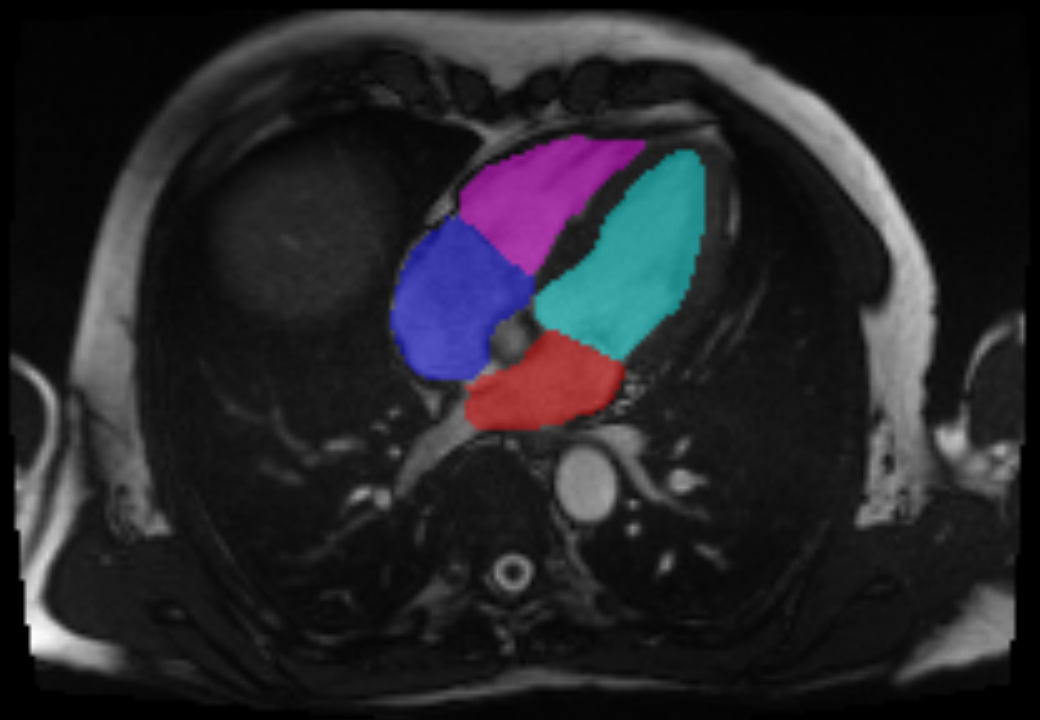}
      \endminipage\hfill
      \minipage{0.33\textwidth}
        \includegraphics[width=\textwidth, height=0.7\textwidth]{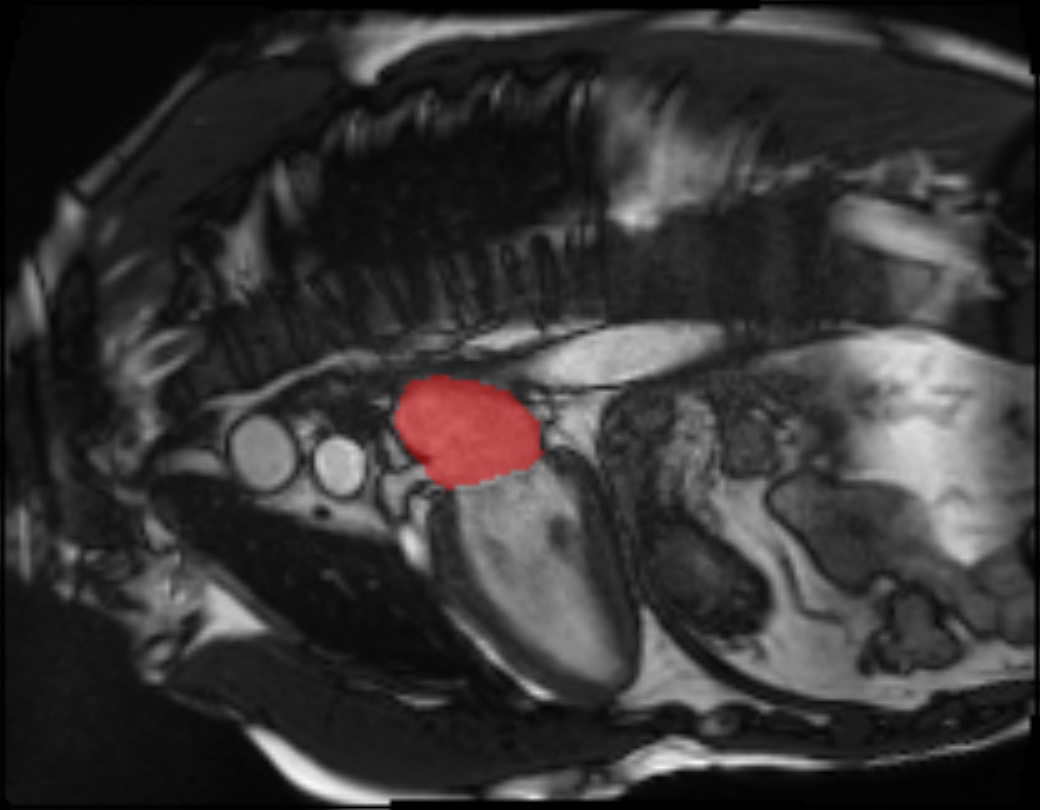}
      \endminipage\hfill
      \minipage{0.33\textwidth}
        \includegraphics[width=\textwidth, height=0.7\textwidth]{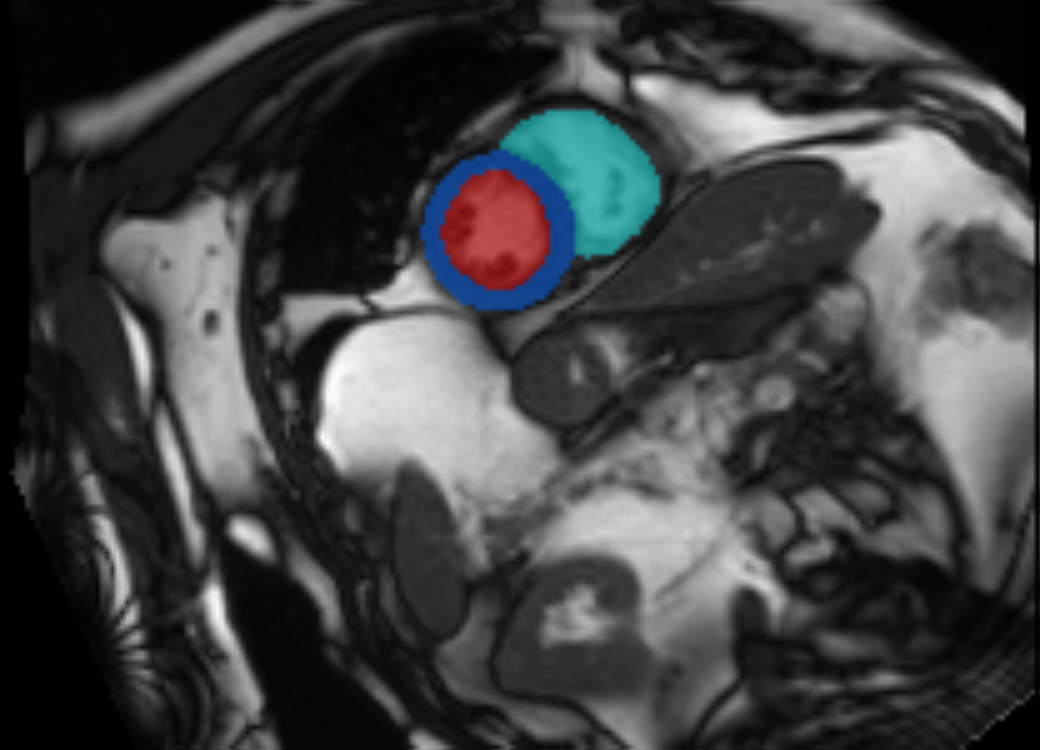}
      \endminipage\hfill
      
      \caption{Example of the segmentation outputs of the long axis 4CH (left), 2CH (middle) CMR view imaging data and the short axis (right) CMR imaging data.}
      \label{fig:example_mri_seg}
      \vspace{-4ex}
    \end{figure}

The CMR imaging data from the UK Biobank that is relevant to MR detection includes long-axis 2-chamber (2CH) view and long-axis 4-chamber (4CH) view, which are all shown in Figure~\ref{fig:example_mri_seg}. In addition, the short-axis view CMR provides accurate description of the left ventricle. Both long-axis views and short-axis view are used to estimate heart measurements relevant to the MR detection task, while only the long-axis 4CH view is used for the deep learning models.

As a pre-processing step, we performed semantic segmentation on the CMR imaging data, using masks (Figure~\ref{fig:example_mri_seg}) generated by a U-Net~\cite{ronneberger2015u} segmentation model to highlight regions of interest to MR classification. U-Net is currently the leading model architecture for medical imaging segmentation, with various U-Net variants developed for different applications. TernausNet~\cite{iglovikov2018ternausnet} is a U-Net variant that reshapes the U-Net encoder to match the VGG11 architecture, allowing it to use pre-trained VGG11~\cite{simonyan2014very} model weights for faster convergence and improved segmentation results. While most medical imaging segmentation models are trained using supervised learning, weakly supervised segmentation methods such as VoxelMorph augmented segmentation~\cite{zhao2019data}, ACNN \cite{oktay2017anatomically}, CCNN \cite{kervadec2019constrained}, graph-based unsupervised segmentation \cite{nian2017graph}, and GAN-based unsupervised segmentation \cite{wu2021unsupervised,wu2020unsupervised} also produce comparable segmentation results. For the segmentation of the 4CH, 2CH, SA, and aorta view CMR imaging dataset from the UK Biobank, Bai et al.~\cite{bai2018automated} offer a supervised segmentation model.

We manually labeled 100 CMR images for each view and trained a supervised segmentation model with the TernausNet~\cite{iglovikov2018ternausnet} architecture. Then, segmentation outputs, shown in Figure~\ref{fig:example_mri_seg}, are used to compute measurements of cardiac structure and function for the four chambers of the heart, as summarized in Table~\ref{tab:cardiac_measurements}. The short-axis view CMR segmentation output is used to estimate the left ventricle and right ventricle measurements, while the long-axis 4CH view and 2CH view outputs are used to estimate the left atrium and right atrium measurements. Specifically, the left atrial volume is estimated using the biplane method with segmentation of both the 2CH and 4CH view, while the right atrial volume is estimated using single plane method with segmentation of the 4CH view.

\vspace{-2ex}
\subsection{Three MR classification models}
\vspace{-2ex}

We consider two baseline models, random forests (Section~\ref{sec:rf}) and a CNN-LSTM (Section~\ref{sec:cnn-lstm}). We then present our CUSSP model in Section~\ref{sec:cussp}.

\vspace{-3ex}
\subsubsection{Random forest baseline}
\label{sec:rf}

We first considered a random forest (RF) classifier \cite{breiman2001random} trained for MR classification on the tabular heart measurements derived from the semantic segmentation masks, as described in Section~\ref{sec:segmentation}. We divided the 18 features by body surface area (BSA) prior to training the RF. 

    \begin{table}

    \vspace{-3ex}
    \centering
    \begin{tabular}{|l|l|l|l|}
    \hline
         Left Atrium & Right Atrium & Left Ventricle & Right Ventricle \\
    \hline
         Vol Max (mL)        & Vol Max (mL)       & End-Systolic Vol (mL) & End-Systolic Vol (mL) \\
         Volume Min (mL)        & Vol Min (mL)       & End-Diastolic Vol (mL) & End-Diastolic Vol (mL) \\
         Stroke Vol (mL)     & Stroke Vol (mL)    & Stroke Vol (mL) & Stroke Vol (mL) \\
         Ejection Fraction (\%) & Ejection Fraction (\%)& Ejection Fraction (\%) & Ejection Fraction (\%) \\
         & & Cardiac Output (L/min) & \\
         & & Mass (g) & \\
    \hline
    \end{tabular}
    \caption{Cardiac measurements derived from the semantic segmentation of the CMR.}
    \label{tab:cardiac_measurements}
    \vspace{-8ex}
    \end{table}

\newpage
\subsubsection{Weakly supervised CNN-LSTM baseline}
\label{sec:cnn-lstm}
\vspace{-3ex}

\paragraph{Conceptualization.} The first deep learning  model for MR classification we developed is a weakly supervised CNN-LSTM following the principles in Fries et al.~\cite{fries2019weakly} and operating on the 4CH CM imaging data. Fries et al.~\cite{fries2019weakly} used CMR imaging sequences from the UK Biobank, however, the objective of their work was the identification of  aortic valve malformations. Their proposed deep learning architecture -- CNN-LSTM -- used DenseNet~\cite{huang2017densely} as the CNN of choice to encode CMR imaging frames and the LSTM to encode embeddings of all frames within each sequence for a final classification of aortic valves into tricuspid (normal) and bicuspid (pathological). We point out that \emph{our MR classification problem is considerably more challenging}. We hypothesize this difficulty is due to the lack of direct view of the mitral valve in the CMR imaging data. Moreover, the flow information provided from the 4CH view CMR imaging data is difficult to learn and encode in the model, an issue which we alleviated in the CUSSP framework.

    \begin{figure}[!ht]
    \vspace{-4ex}
        \includegraphics[width=\textwidth]{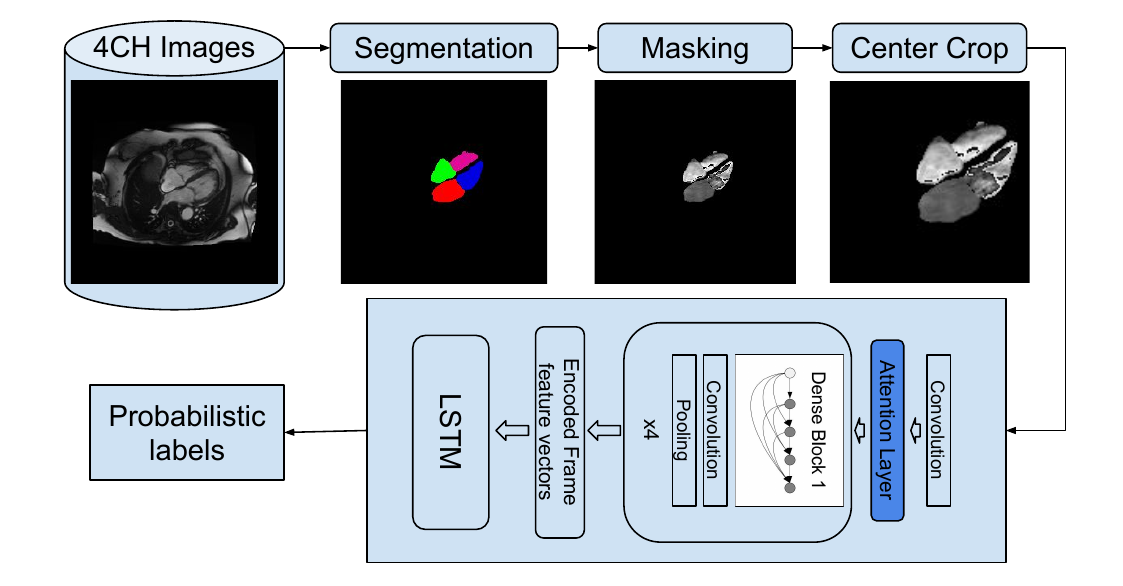}
        \caption{Overview of the CNN-LSTM method pipeline for MR classification.}
        \label{fig:CNN-LSTM-diagram}
        \vspace{-6ex}
    \end{figure}
    
\paragraph{Test-time pipeline.} 
The CNN-LSTM pipeline, shown in Figure~\ref{fig:CNN-LSTM-diagram} includes an image segmentation model, and an image classification model. It uses the 4CH CMR from the UK Biobank. The CMR data is center-cropped using the center of mass of the CMR imaging frames. The resulting sequence provided to the CNN-LSTM, which generates probabilistic labels of MR for the sample.

\paragraph{Training process.}

In the CNN-LSTM model architecture, the CNN serves as the frame encoder, which encodes each frame of each sequence into a representation vector. The model uses DenseNet-121 pre-trained on ImageNet as the CNN. To better learn the attention span of the frame encoder, we added an attention layer to the DenseNet-121 after the first convolutional layer. After the bi-directional LSTM, a multi-layer perceptron (MLP) performs the final classification.

\subsubsection{The CUSSP framework}
\label{sec:cussp}

\paragraph{Conceptualization.} To better encode the blood flow information relevant to MR classification from the 4CH CMR view, we investigated self-supervised representation learning methods which can leverage all the unlabeled CMR sequences present in the UK Biobank. 
Typically, self-supervised representation learning for visual data involves maximizing the similarity between representations of various distorted versions of a sample. Among the many self-supervised architectures, SimCLR~\cite{chen2020simple}, SwAV~\cite{caron2020unsupervised}, and BYOL~\cite{grill2020bootstrap}, we chose Barlow Twins~\cite{zbontar2021barlow},
since it does not require large batches.
With the labeled data, our siamese network compares the representation differences between classes by sampling two inputs from different classes as performed in~\cite{xing2018offline}. Thus, our CUSSP MR classification pipeline takes advantage of both self-supervised and supervised representation learning.

\paragraph{Test-time pipeline.} 

\begin{figure}[!ht]
\vspace{-4ex}
    \includegraphics[width=\textwidth]{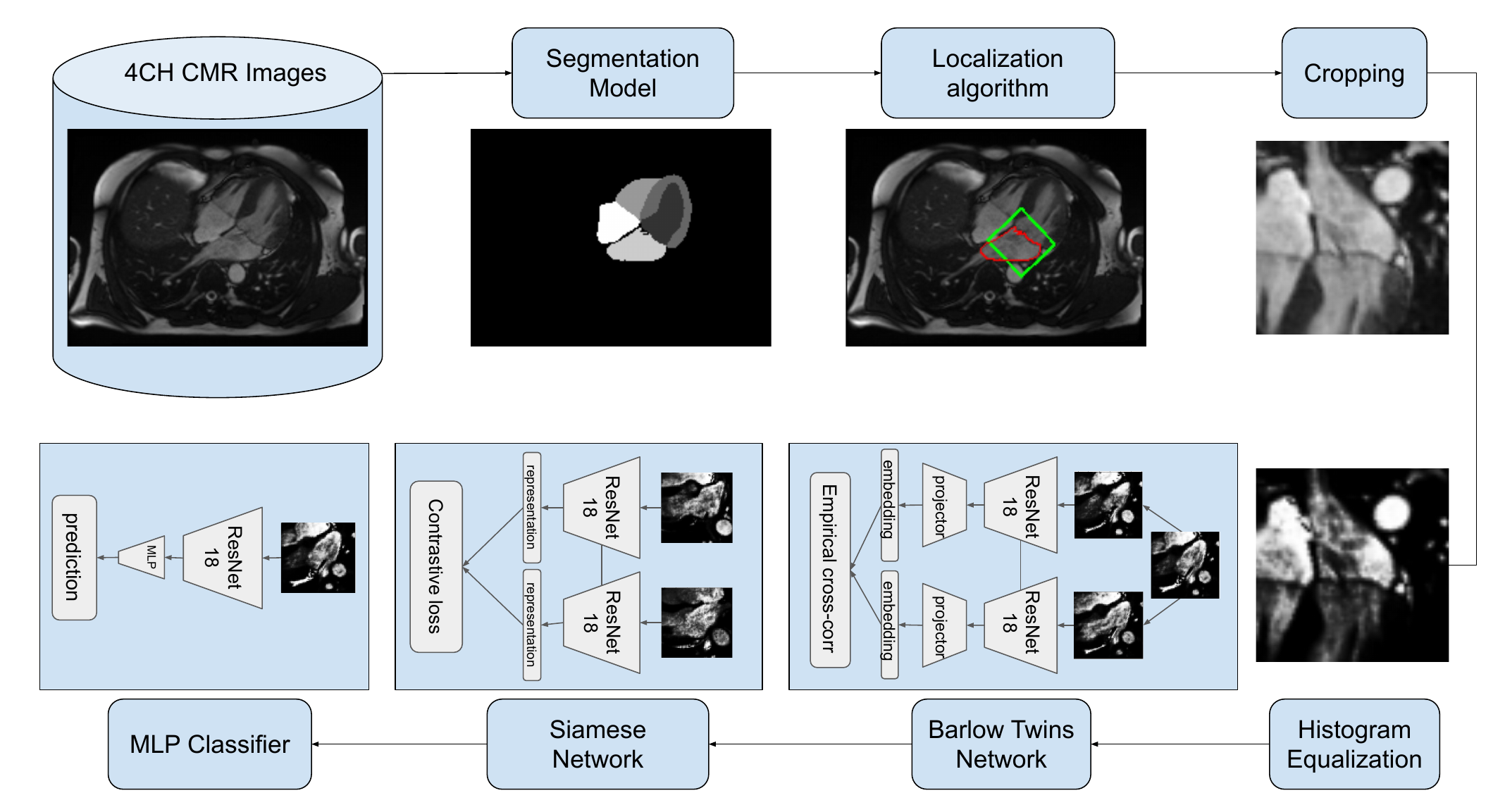}
    \caption{Overview of the CUSSP pipeline for MR classification, with its 5 steps: (1) segmentation, (2) localization, (3) cropping, (4) equalization, and (5) prediction.}
    \label{fig:mr-btsm-pipeline}
    \vspace{-6ex}
\end{figure}

Our CUSSP method consists of five main steps, shown in Figure~\ref{fig:mr-btsm-pipeline}, with the first two steps representing data preprocessing, and the later three steps using network components trained for MR classification, as described in the next section. The pre-processing of the CMR imaging sequence is shown in Figure~\ref{fig:barlow-data-prep} in the Appendix. We used the segmentation model in~\ref{sec:segmentation} to locate the mitral valve and the orientation of the left ventricle. We then cropped a square patch with the mitral valve at its center positioned horizontally. After cropping, we applied histogram equalization to the patch with the pixel intensity range of the left atrium. The resulting patches are used by the downstream networks.

\paragraph{Training process.}
The first step involves training a representation encoder in a Barlow Twins network using over 30,000 unlabeled pre-processed sequences. ResNet-18 with an output dimension of 512 is used as the encoder, with hidden dimension and the projector output dimensions being 2048. After training the encoder with the unlabeled dataset, it is fine-tuned in a siamese network using a comparatively smaller labeled set, as indicated in~\ref{sec:exp_setup}. During training, two sequences are sampled from the labeled dataset, with the first being non-MR and the second being either MR or non-MR. The two sequences are passed through the representation encoder to obtain embeddings, which are then used to calculate the contrastive loss. The model is trained to maximize contrastive loss when the two samples are non-MR and MR and to minimize it when both are non-MR. Once the representation encoder is fine-tuned in the siamese network, it is combined with a 3-layer multi-layer perceptron (MLP) network to form a classifier, which is trained on the same labeled dataset. To improve computation efficiency and training accuracy, we also tested the framework using a smaller window of 25 frames, since MR occurs between diastole and systole.

\section{Experiments}

\subsection{Experimental setup}
\label{sec:exp_setup}

\noindent \emph{Data splitting.} 4CH CMR images were used to conduct experiments with both the CNN-LSTM method and the CUSSP method. We used a total of 704 labeled sequences, with 525 sequences selected for the training set, including 452 labeled as non-MR and 73 labeled as MR. The remaining 179 sequences were used for testing, with 154 labeled as non-MR and 25 labeled as MR. 

\noindent \emph{Evaluation metrics.} Considering the substantial class imbalance, we opted to use F1 score as our primary evaluation metric, along with precision and recall.

\subsection{Random Forest Classification Results}
The random forest model is trained with 10-fold cross validation, with a random search over a parameter grid of $n\_estimators (10-100)$, $max\_depth (2-16) $, $max\_features (sqrt, log2)$, $min\_samples\_leaf (2-8) $. The optimal hyper-parameter setting found is: $n\_estimators=20$, $max\_features=log2$, $max\_depth=8$, $min\_samples\_leaf=2$. The best results obtained are presented in Table \ref{table:complete-results}.
    
    \begin{table}

    \vspace{-4ex}
    \centering
    \begin{tabular}{|l|l|l|l|l|l|l|}
    \hline
        Model     &    Pos. Acc  &  Neg. Acc & Precision & Recall & F1    & AUC        \\
    \hline
    \hline
        RF   &    0.09     &  0.99      & 0.43       & 0.09  & 0.14   & 0.58   \\
    \hline
    CNN-LSTM &    0.53     &  0.86    & 0.45     & 0.53  & 0.44    &   0.72   \\
    \hline
   CUSSP-1 &    0.38     &  0.87    & 0.29     & 0.38  & 0.32 & 0.65 \\
   \hline
   CUSSP-2 &    0.29     &  0.87    & 0.25     & 0.29  & 0.27 & 0.63 \\
   \hline
   CUSSP-3 &    0.38     &  0.90    & 0.35     & 0.38  & 0.36 & 0.66 \\
   \hline
   CUSSP-SIAM  &    0.55     &  0.96    & 0.66     & 0.55  & 0.60 & 0.80 \\
   \hline
   CUSSP-SIAM-25 \qquad &    \textbf{0.62}     &  \textbf{0.96}    & \textbf{0.8}     & \textbf{0.62}  & \textbf{0.69} & \textbf{0.88} \\
    \hline
    \end{tabular}
    
    \caption{Experimental results for Random Forest (RF) baseline, CNN-LSTM and CUSSP. CUSSP-1, CUSSP-2 and CUSSP-3 are trained with the BarlowTwins-MLP model without fine-tuning with the Siamese network. CUSSP-SIAM and CUSSP-SIAM-25 are trained with the BarlowTwins-Siamese-MLP model.}
    \label{table:complete-results}

    \vspace{-10ex}
    \end{table}

\subsection{CNN-LSTM Classification Results}

    We conducted experiments on the DenseNet-LSTM classification model using various input image sizes, attention layer configurations, and masks. The best CNN-LSTM model attains a F1-score of 0.44, shown in Table \ref{table:complete-results}, with further information on the performance under other settings summarized in the Appendix.

\subsection{CUSSP Classification Results}
    
    We evaluated various configurations of the CUSSP model, to determine the relative benefits of different components. In the first configuration, the ResNet18 model was combined with a 3-layer MLP to train a classifier using the labeled training set after being trained in the Barlow-Twins network with the unlabeled dataset. During the classifier training, the cross-correlation loss from the Barlow-Twins network and the cross-entropy loss from the binary classification were weighted using three different configurations. For CUSSP-1 the cross-correlation loss has a weight of 0.9, while the cross-entropy loss has a weight of 0.1. For CUSSP-2, the weights are 0.5 and 0.5, while for CUSSP-3 they are 0.1 and 0.9, respectively. Both CUSSP-1 and CUSSP-3 outperform CUSSP-2, though the performance is low, indicating the importance of fine-tuning, described below.
    
\begin{figure}[!ht]
    \vspace{-4ex}
    \centering
      \includegraphics[width=0.9\textwidth]{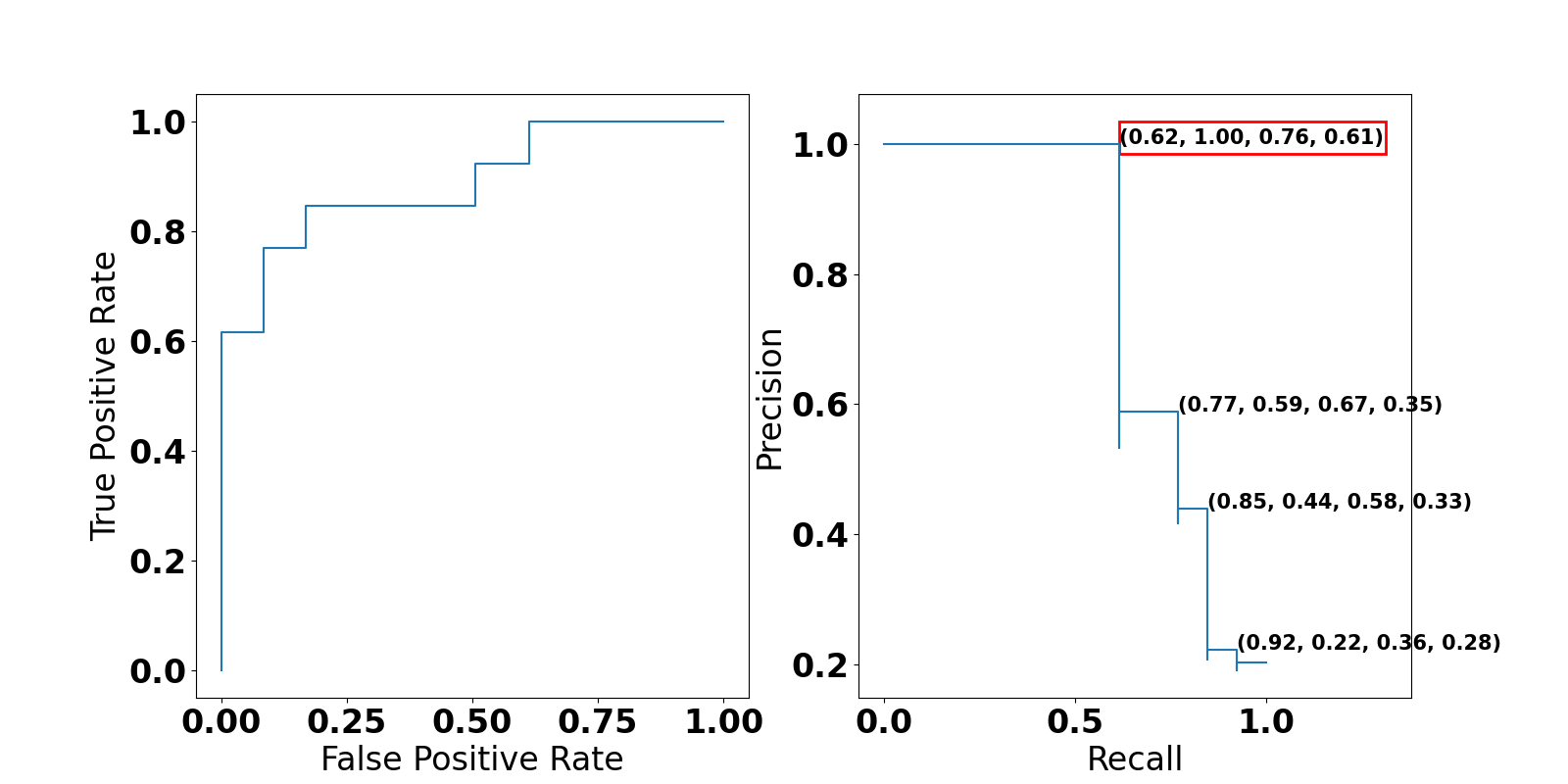}
      \caption{The ROC AUC curve and the precision-recall curve of CUSSP. The annotated coordinates on the precision-recall curve plot are (recall, precision, f1-score, threshold).}
      \label{fig:barlow-mlp-curve}
      \vspace{-7ex}
    \end{figure}

    In the second scenario, we fine-tuned the encoder with a siamese network to enhance the quality of the encoded representations after training the Barlow Twins network. To prevent overfitting of the model and to limit its capacity, we froze the parameters of all layers except the last block of the ResNet18 encoder when training the siamese network and the classifier. The resulting model, CUSSP-SIAM, showed a significant improvement in performance. In the final configuration CUSSP-SIAM-25, the number of frames in the training sequences was truncated from 50 frames to the 25 frames that correspond to the interval when mitral regurgitation occurs. The results are summarized in Table~\ref{table:complete-results}, while the ROC-AUC curve for CUSSP-SIAM-25 are shown in Figure~\ref{fig:barlow-mlp-curve}.



\section{Conclusion}

We present the world's first automated mitral regurgitation classification system. The CUSSP model we developed, trained with limited supervision, operates on 4CH CMR imaging sequences and attains an F1 score of 0.69 and an ROC AUC of 0.88, opening up the opportunity for large-scale screening for MR.

\bibliographystyle{splncs04}
\bibliography{mybibliography}

%





\newpage
\section*{Appendix}

   \begin{figure}
      \minipage{0.33\textwidth}
        \includegraphics[width=\textwidth, height=0.7\textwidth]{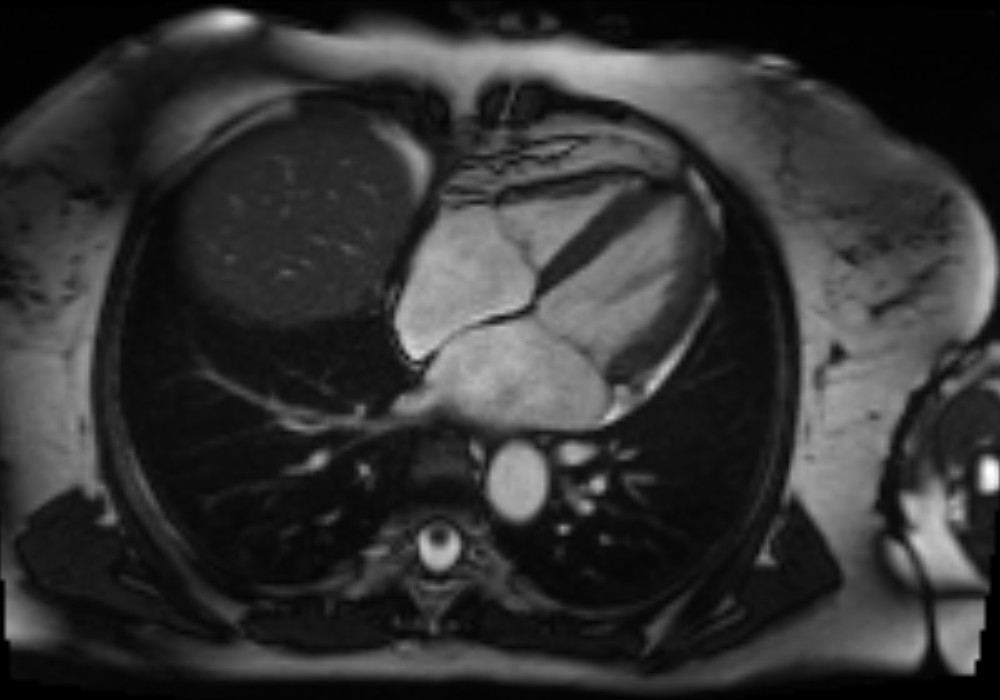}
      \endminipage\hfill
      \minipage{0.33\textwidth}
        \includegraphics[width=\textwidth, height=0.7\textwidth]{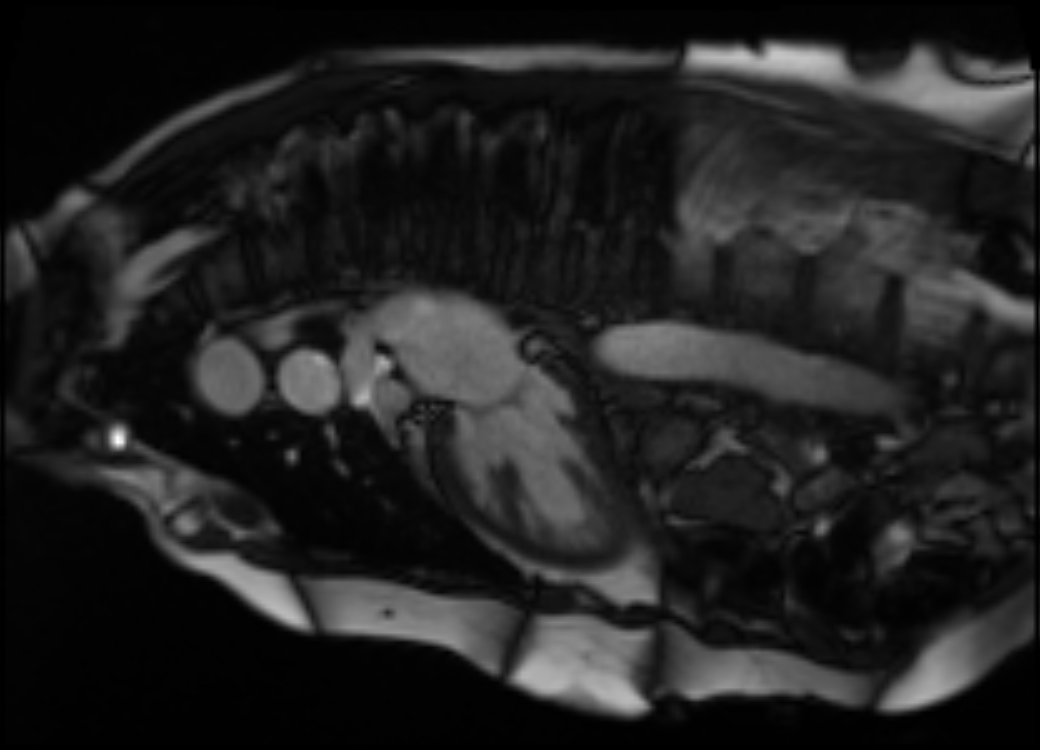}
      \endminipage\hfill
      \minipage{0.33\textwidth}
        \includegraphics[width=\textwidth, height=0.7\textwidth]{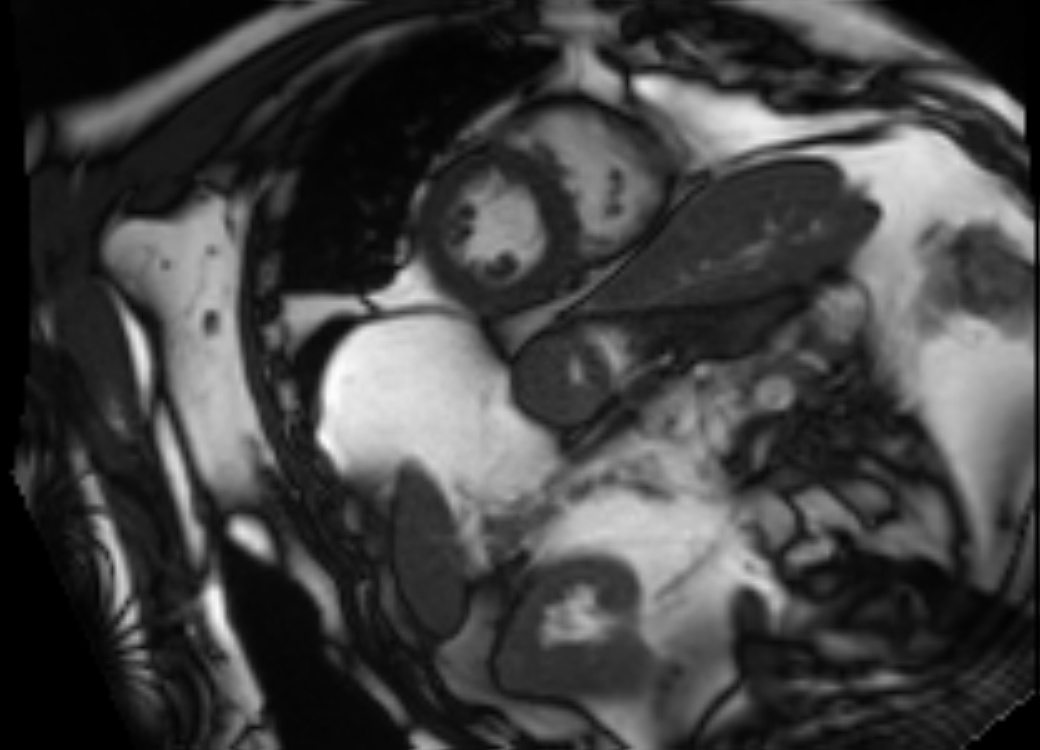}
      \endminipage\hfill

      \minipage{0.33\textwidth}
        \includegraphics[width=\textwidth, height=0.7\textwidth]{figs/example_4Ch_seg.png}
      \endminipage\hfill
      \minipage{0.33\textwidth}
        \includegraphics[width=\textwidth, height=0.7\textwidth]{figs/example_2Ch_seg.png}
      \endminipage\hfill
      \minipage{0.33\textwidth}
        \includegraphics[width=\textwidth, height=0.7\textwidth]{figs/example_sa_seg.png}
      \endminipage\hfill
      
      \caption{Overview of the dataset. Top: Example of the long axis 4CH (left), 2CH (middle) CMR view imaging data and the short axis (right) CMR imaging data. Bottom: Example of the segmentation outputs of the long axis 4CH (left), 2CH (middle) CMR view imaging data and the short axis (right) CMR imaging data.}
      
      \label{fig:example_mri-with-segmentation}
    \end{figure}

    \vspace{-5ex}

    \begin{figure}[!ht]
        \includegraphics[width=\textwidth]{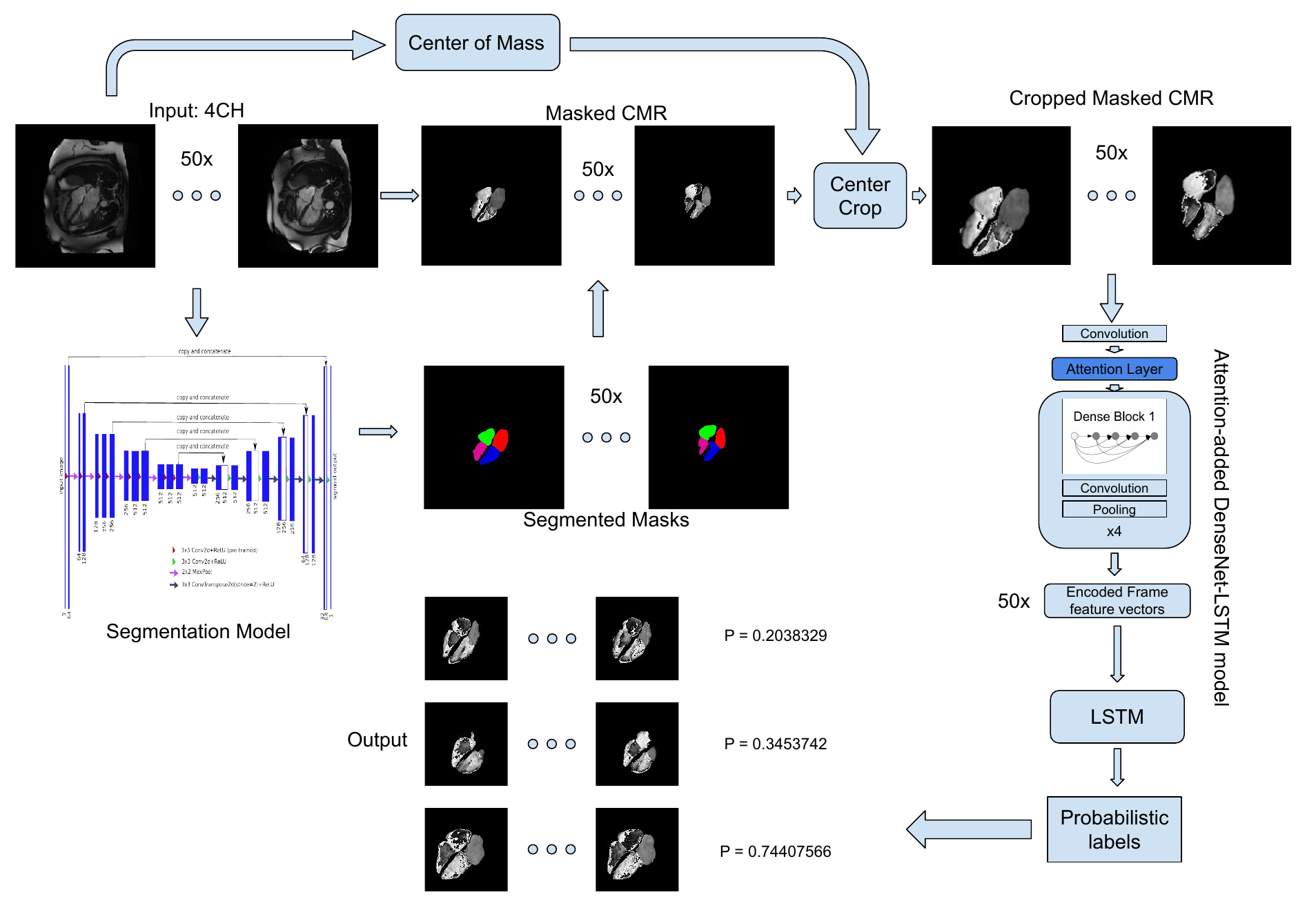}
        \caption{Detailed overview of the CNN-LSTM method pipeline for MR classification.}
        \label{fig:CNN-LSTM-diagram-detailed}
    \end{figure}

    \begin{figure}
        \includegraphics[width=.98\textwidth, height=.48\textwidth]{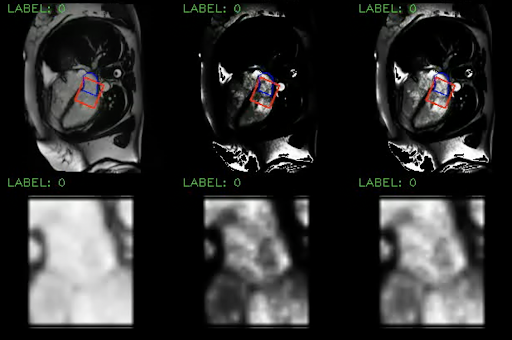}
        \caption{Detailed overview of the pre-processing steps for CUSSP. Top: Example of the 4CH CMR images in the original contrast (left), the left atrium histogram equalized contrast (middle), and the cropped patch histogram equalized contrast (right), with blue contours outline the left atrium, and the red square boxes outline the patch to crop. Bottom: Example of the cropped mitral valve patch as outlined in the red square boxes in the top row.}
        \label{fig:barlow-data-prep}
    \end{figure}



 \begin{figure}[!ht]
     \includegraphics[width=.98\textwidth, height=.48\textwidth]{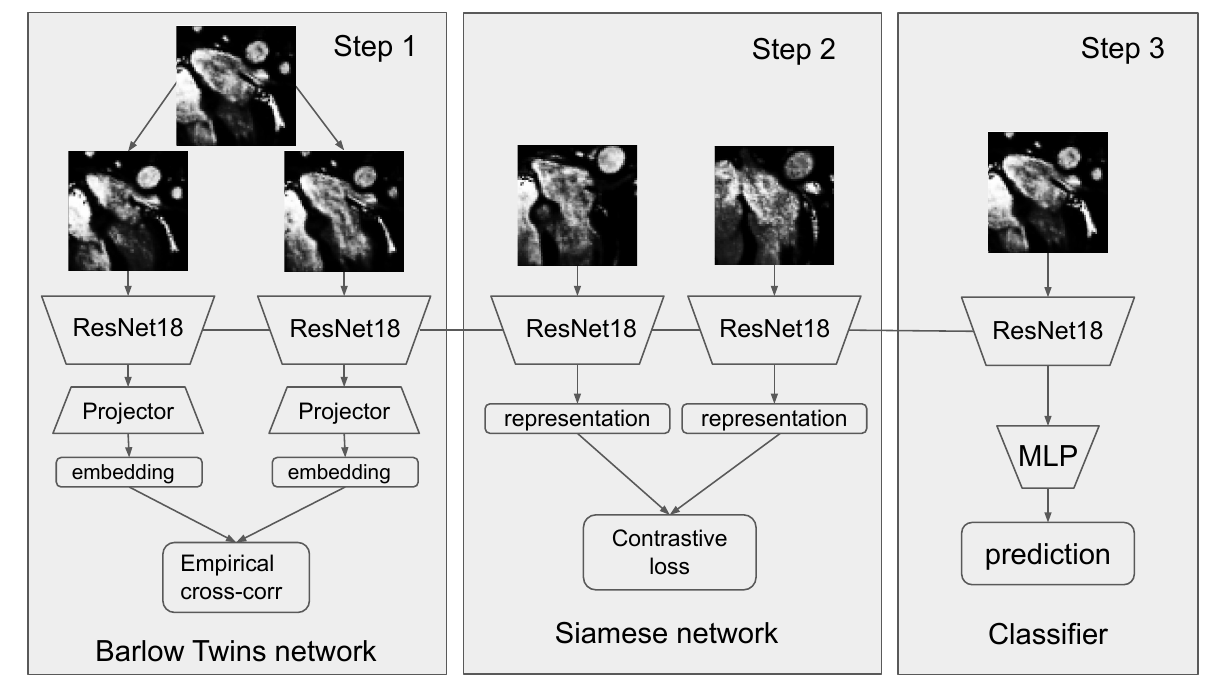}
     \caption{The model training stage of the CUSSP method contains three steps: (i) the feature encoder is trained in the Barlow-Twins network with unlabeled imaging data set, (ii) the feature encoder is fine-tuned in a siamese network with labeled imaging data set, and (iii) the feature encoder is assembled with a MLP, then trained with labeled imaging data set for the classification task of MR.}
     \label{fig:btsm-pipeline}
 \end{figure}

\end{document}